\documentclass[a4paper,conference]{IEEEtran}

\usepackage{array}
\newcolumntype{C}[1]{>{\centering\let\newline\\\arraybackslash\hspace{0pt}}m{#1}}
\usepackage{cite} 

\usepackage{graphicx}  

\usepackage{subfig}

\usepackage{float} 

\usepackage{url}       

\usepackage{comment}
\usepackage{stfloats}  

\usepackage{amsmath}   
\usepackage{fancyheadings}
\begin{document}

\title{SPH Modeling of Short-crested Waves}

\author{\IEEEauthorblockN{Zhangping Wei}
\IEEEauthorblockA{Department of Civil Engineering\\
Johns Hopkins University\\
Baltimore, MD 21218, USA\\
zwei@jhu.edu; zwei.coast@gmail.com}
\and
\IEEEauthorblockN{Robert A. Dalrymple}
\IEEEauthorblockA{Department of Civil Engineering\\
Johns Hopkins University\\
Baltimore, MD 21218, USA\\
rad@jhu.edu}}

\maketitle

\begin{abstract}
This study investigates short-crested wave breaking over a planar beach by using the mesh-free Smoothed Particle Hydrodynamics model, GPUSPH. The short-crested waves are created by generating intersecting wave trains in a numerical wave basin. We examine the influence of beach slope, incident wave height, and incident wave angle on the generated short-crested waves. Short-crested wave breaking over a steeper beach generates stronger rip currents, and larger circulation cells in front of the beach.
Intersecting wave trains with a larger incident wave height drive a more complicated short-crested wave field including isolated breakers and wave amplitude diffraction.
Nearshore circulation induced by short-crested wave breaking is greatly influenced by the incident wave angle (or the rip current spacing). There is no secondary circulation cell between the nodal line and the antinodal line if the rip current spacing is narrow. However, there are multiple secondary circulation cells observed when the rip current spacing is relatively large.

\end{abstract}

\section{Introduction}
Although there are considerable amount of Smoothed Particle Hydrodynamics (SPH) applications in the field of coastal engineering, most of them address wave-structure interaction (e.g., \cite{wei2016numerical}, \cite{wei2015sph}, and \cite{wei2016simulation}, among many others) and a very few concern surf zone waves.
By use of the open-source Smoothed Particle Hydrodynamics model GPUSPH \cite{herault2010sph}, we recently conducted numerical experiments to investigate short-crested waves in the surf zone. In Wei et al.\cite{wei2017short}, we generated short-crested waves by superimposing intersecting wave trains in a numerical wave basin and examined short-crested wave breaking over a beach. Due to the relatively strong nonlinearity, the generated short-crested waves breaking at the toe of the planar beach, and we observed rip currents and undertow. We also found that the superposition of intersecting waves and the interaction between rip current and the crest end of short-crested waves create smaller isolated breakers. Furthermore, wave amplitude diffraction at these isolated waves gives rise to an increase in the alongshore wave number in the inner surf zone. We also observed the existence of 3D vortices and multiple circulation cells with a rotation frequency much lower than the incident wave frequency. We also measured the vertical vorticity, an indicator of horizontal rotation of flows, generated by short-crested wave breaking over the beach. 

Although our observations in \cite{wei2017short} are interesting and significant, they were based on one realization of complicated surf zone waves, and more effort is needed to examine surf zone processes with different conditions. In Wei and Dalrymple\cite{wei2017sph}, we re-examined the wave phenomena observed in Wei et al.\cite{wei2017short} by considering a smaller incident wave height. Short-crested waves generated by a smaller incident wave height break over the upper planar beach, resulting in a weaker rip current field. We obtained different nearshore circulation pattern and vertical vorticity field when comparing with those created by a higher incident wave. In addition to incident wave height, many other factors also influence surf zone waves. This observation motivates us to examine some of them by conducting extra numerical experiments in this study.

The rest of the paper is organized as follows. We first briefly review the generation of short-crested waves by using the synchronous wave trains method of Dalrymple~\cite{dalrymple1975mechanism} in section~\ref{governing_equations}. Section~\ref{exp_setup} presents the numerical experiment setup for examining incident wave height, incident wave angle, and beach slope on the generated short-crested waves. Sections~\ref{wave_profile}, \ref{current_field}, and \ref{circulation} present the breaking wave profile, current field, and vertical vorticity field, respectively. Finally we summarize the findings of this work in section \ref{conclusions}. 

\section{The Synchronous Wave Trains Method}\label{governing_equations}
This study generates short-crested waves by following Dalrymple~\cite{dalrymple1975mechanism} by superimposing intersecting wave trains of the same wave period. The detailed derivation for wave generation has been given by Wei et al.~\cite{wei2017short}, and we review it briefly for the completeness of this work. Given a coordinate system ($x$, $y$, $z$), where position $x$ axis is the onshore direction, and $y$ axis is the alongshore direction, we assume that two wave trains of the same period $T$ with the same amplitude propagate to the shore from two different directions, such that the wave rays make angles $\alpha$ and $\beta$ clockwise from the positive $x$ axis. The free surface profile at the offshore boundary (i.e., $x$ = 0) is obtained by
\begin{equation}
\eta (0, y, t) = \frac{H}{2}  \cos \left [ \frac{k}{2}(\sin \alpha_w + \sin \beta_w)y+\sigma t\right ] \cos \left ( \frac{\pi}{\lambda }y\right ) \label{eq:eta4}
\end{equation}
where $t$ is the time; $k$ is the wave number ($2\pi/L$);  $L$ is the wave length; $\sigma$ is the angular frequency ($2\pi/T$); $H$ is the wave height after superimposing the intersecting wave trains; the subscript $w$ indicates information at wavemakers. The $\lambda$ term in Eq.~(\ref{eq:eta4}) is the spacing between rip currents or the distance between two nodal (or antinodal) lines. It is defined as 
\begin{equation}
\lambda =\frac{L}{\sin \alpha_w - \sin \beta_w } \label{eq:lambda2}
\end{equation}

\begin{figure}
\centering
  \includegraphics[width=0.45\textwidth]{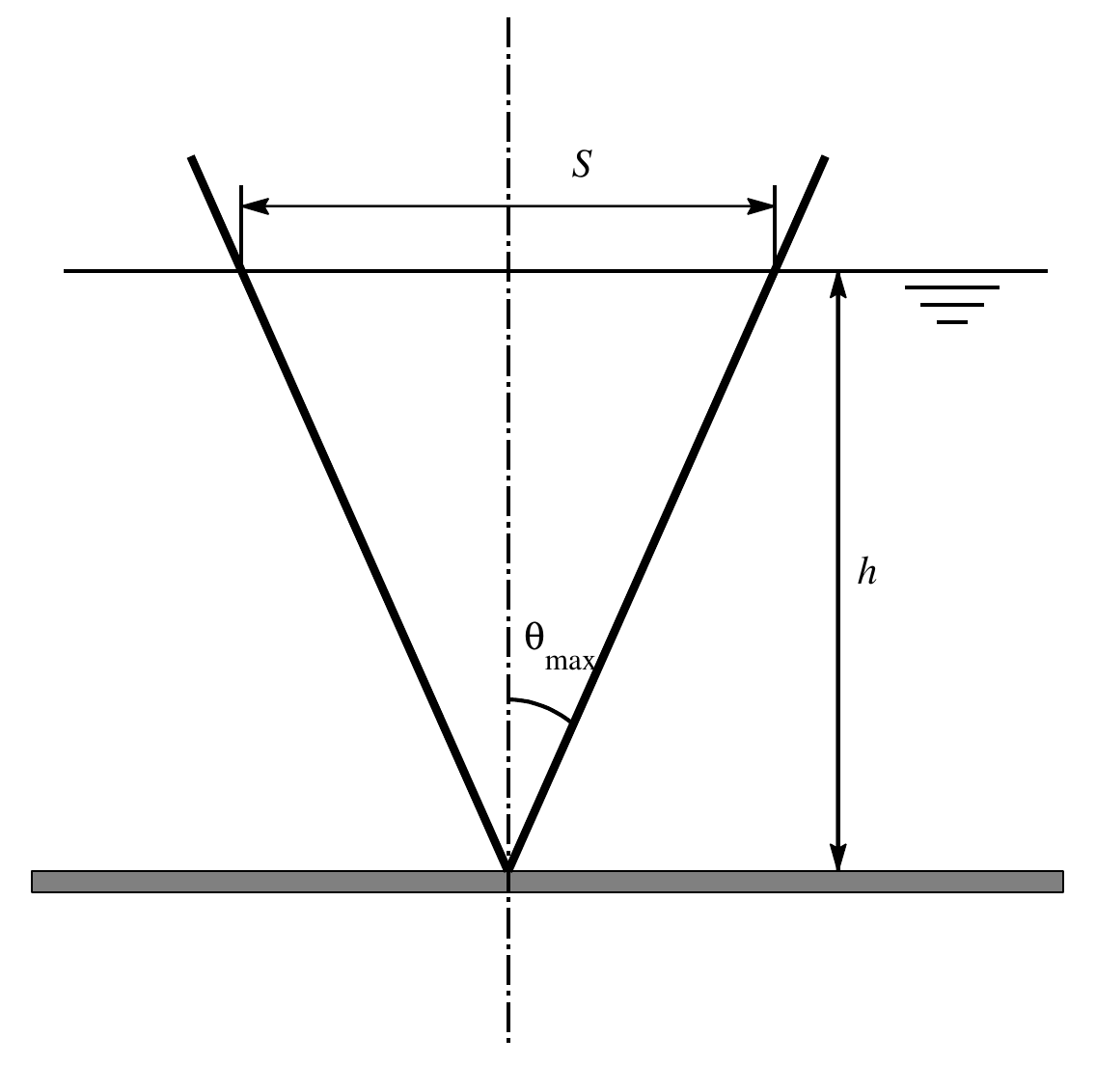}
\caption{Definition sketch of a flap-type wavemaker \cite{wei2017short}. }
 \label{fig:setup}
\end{figure} 

Considering a flap-type wavemaker as shown in Fig.~\ref{fig:setup}, the maximum rotation angle is defined by

\begin{equation}
\theta_{max} =\tan^{-1} \left (\frac{S/2}{h}\right )\label{eq:angle1}
\end{equation}
where $h$ is the local water depth at the flap-type wavemaker, and the stroke $S$ is given by Dean and Dalrymple \cite{dean1991water} as

\begin{equation}
\frac{H}{S}=4\left ( \frac{\sinh kh}{kh} \right )\frac{kh\sinh kh -\cosh kh + 1}{\sinh 2kh + 2kh}\label{eq:HS}
\end{equation}

To generate short-crested waves with the alongshore variation of free surface profile as indicated in Eq.~(\ref{eq:eta4}), a ``snake'' wavemaker that consists of a series of flap-type wavemakers is used in this study. Correspondingly, the instantaneous rotation angle of individual flap-type wavemaker is determined by

\begin{equation}
\theta (0, y, t) = \theta_{max} \cos \left [ \frac{k}{2}(\sin \alpha_w + \sin \beta_w)y+\sigma t\right ] \cos \left ( \frac{\pi}{\lambda }y\right )\label{eq:angle4}
\end{equation}

\section{Numerical Experiment Setup}\label{exp_setup}
The capability of GPUSPH to generate and predict short-crested waves has been verified by comparing with laboratory experiments in \cite{wei2017sph} and \cite{wei2017short}. This study uses a numerical wave basin similar to the one used in our previous studies. For example, we still choose the alongshore width of the basin to be 20~m; the length of the offshore horizontal flat is $b_x$ = 4~m; the offshore water depth is set to be $h$ = 0.5 m; and a series of flap-type wavemakers are located along the $y$ axis at $x$ = 1~m. Furthermore, we also consider two incident wave trains with the same wave period of $T = 2$~s and equal but opposite incident wave angles. Regarding the numerical discretization in GPUSPH, the full basin is discretized into particles, with a fixed particle size of $\Delta p$ = 0.02~m, resulting in 25 particles over the water column offshore. At the two alongshore boundaries at the antinodal lines, the no-flow wall boundary condition is applied.

To address the impact of incident wave angle, beach slope, and incident wave height on surf zone waves, we consider four numerical experiments, as listed in Table~\ref{tab:wave}. It is worth pointing out that since all cases have the same offshore water depth and wave period, they will have the same cross-shore wave length ($L$). As a result, the rip current spacing in Eq.~(\ref{eq:lambda2}) is solely determined by incident wave angles. The breaking wave type is estimated by using the Iribarren number\cite{battjes1975surf}
\begin{equation}
\xi_b = \frac{\tan \phi }{\sqrt{H_b/L_0}}
\end{equation}
where $\phi$ is the beach slope; $H_b$ is the value of the wave height at the break point (the incident wave height was used to approximate it in this study); and $L_0$ is the deep-water wave length. The approximate Iribarren number indicates a spilling breaker for all cases. Among these four cases, \textit{Cases} $\bf{a}$ and $\bf{d}$ were reported in Wei and Dalrymple\cite{wei2017sph} and Wei et al.\cite{wei2017short}, respectively. We treat \textit{Case} $\bf{a}$ as the \textit{Baseline} case, and design the other three cases based on it. Specifically, \textit{Case} $\bf{b}$ has the same incident wave height and beach slope as \textit{Case} $\bf{a}$, but with a larger incident wave angle (or a narrower rip current spacing); \textit{Case} $\bf{c}$ has the same incident wave height and rip current spacing as \textit{Case} $\bf{a}$, but with a larger beach slope; and \textit{Case} $\bf{d}$ has the same beach slope and rip current spacing as \textit{Case} $\bf{a}$, but with a larger incident wave height. 

\begin{table}[]
\tiny
\centering
\renewcommand{\arraystretch}{1.3}
\caption{Numerical parameters used to generate short-crested waves in GPUSPH. $H$ is the wave height; $\phi$ is the beach slope; $\lambda$ is the rip current spacing; and $\xi_b$ is the Iribarren number.}
\label{tab:wave}
\resizebox{0.475\textwidth}{!}{%
\begin{tabular}{|c|c|c|c|c|}
\hline
Case No. & $H$ (m) & \multicolumn{1}{l|}{$\phi$} & \multicolumn{1}{l|}{$\lambda$ (m)} & \multicolumn{1}{l|}{\bf{$\xi_b$}} \\ \hline
\bf{a} & 0.2 & 0.02 & 10 & 0.11\\ \hline
\bf{b} & 0.2 & 0.02 & 5 & 0.11\\ \hline
\bf{c} & 0.2 & 0.04 & 10 & 0.22 \\ \hline
\bf{d} & 0.3 & 0.02 & 10 & 0.09 \\ \hline
\end{tabular}%
}
\end{table}

\begin{figure*}[!t]
\centering
  \includegraphics[width=0.9\textwidth]{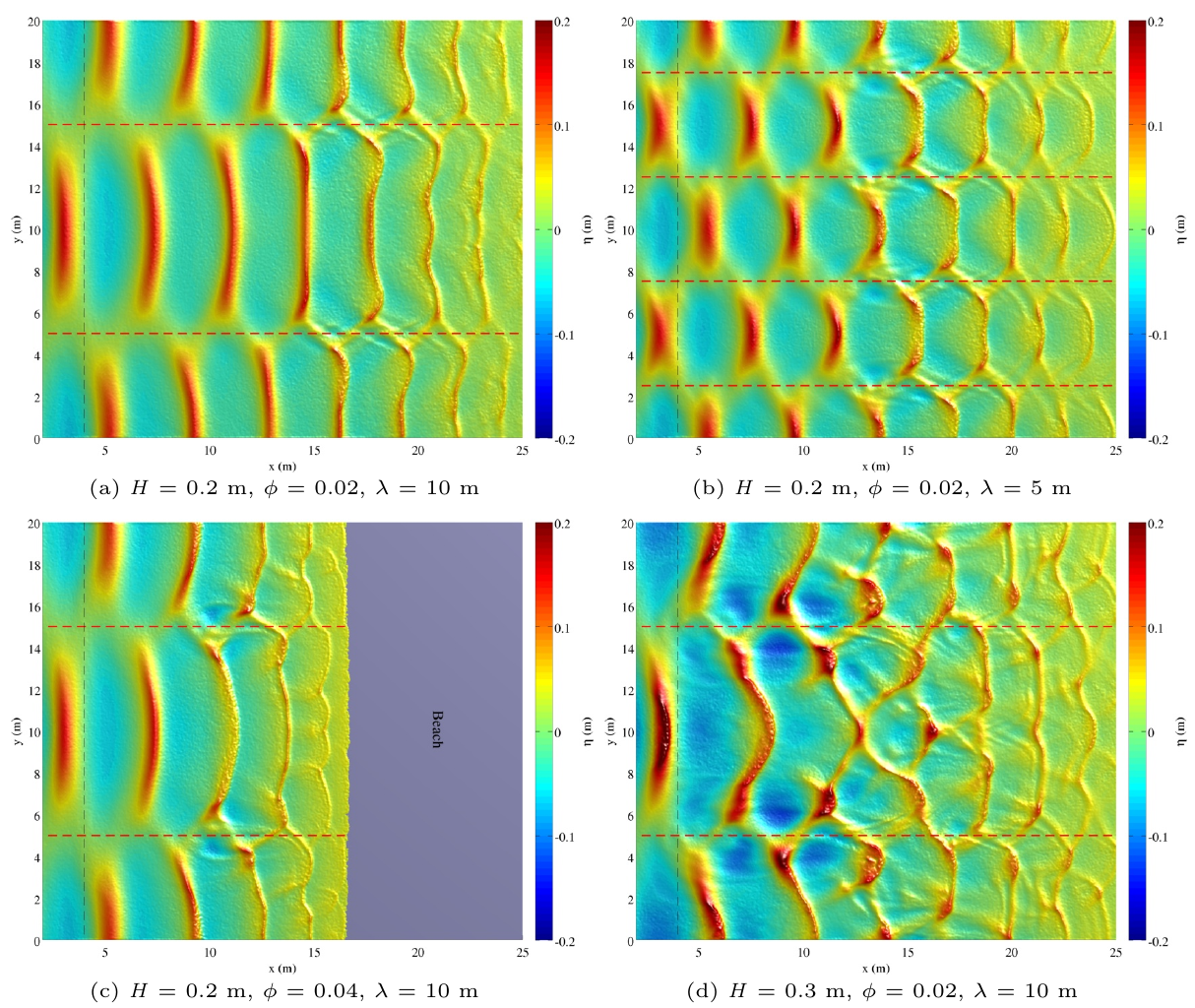}
\caption{The free surface profile of short-crested wave breaking over a beach among different numerical experiments. The dashed line at $x$ = 4~m indicates the starting point of the beach, and the red dashed lines indicate the nodal lines. $H$ indicates the wave height; $\phi$ indicates the beach slope; $\lambda$ indicates the rip current spacing.}
 \label{fig:eta}
\end{figure*}
\section{Breaking Wave Profile}\label{wave_profile}
\begin{figure*}[!t]
\centering
  \includegraphics[width=0.75\textwidth]{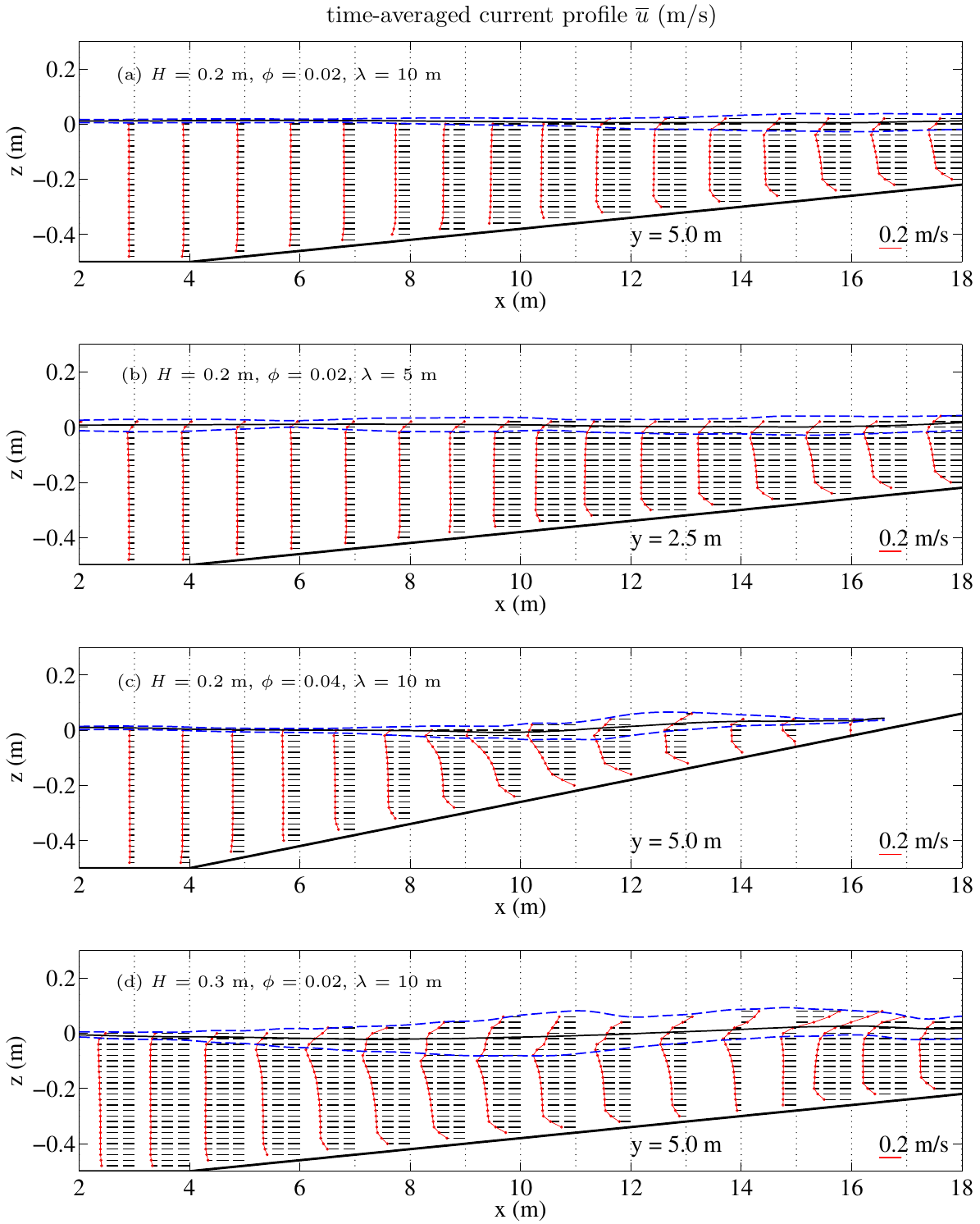}
\caption{The vertical profile of the time-averaged cross-shore velocity at the nodal line (i.e., the location of rip current) among different numerical experiments. The results are based on $n_w$ = 10 waves averaging from $t$ = 80 to 100~s. The zero velocity is set at the individual cross-shore location (e.g., $x$ = 10~m); the plot is at its right side if the velocity is positive, otherwise the plot is at its left side. The black solid line indicates the mean water level; the two blue dashed lines above and below the mean water level indicate the wave crest and the wave trough, respectively; the vertical resolution is one particle size of $\Delta p$ = 0.02~m. $H$ indicates the wave height; $\phi$ indicates the beach slope; $\lambda$ indicates the rip current spacing.}
 \label{fig:vertical_current}
\end{figure*}
Fig.~\ref{fig:eta} compares the free surface profile of short-crested wave breaking over a beach among different numerical experiments as described in the previous section. For \textit{Case} $\bf{a}$ in Fig.~\ref{fig:eta}(a), it has a relatively wide wave crest confined by two nodal lines at $y$ = 5 and 15~m. Waves are subject to shoaling once they propagate over the planar beach starting at $x$ = 4~m, resulting in a curvy wave crest. Eventually waves break around $x$ = 17~m and then propagate nearshore as bores. For \textit{Case} $\bf{b}$ in Fig.~\ref{fig:eta}(b), the rip current spacing is equal to one half of the one in \textit{Case} $\bf{a}$, resulting in a shorter wave crest and four nodal lines at $y$ = 2.5, 7.5, 12.5, and 17.5~m. It is also seen that waves in \textit{Case} $\bf{b}$ break slightly earlier than waves in \textit{Case} $\bf{a}$ around $x$ = 16~m. For \textit{Case} $\bf{c}$ in Fig.~\ref{fig:eta}(c), it has a steeper beach than the one in \textit{Case} $\bf{a}$. Waves in \textit{Case} $\bf{c}$ also break earlier than waves in \textit{Case} $\bf{a}$ around $x$ = 10~m. For \textit{Case} $\bf{d}$ in Fig.~\ref{fig:eta}(d), it has a wave height that is 50\% higher than the one in \textit{Case} $\bf{a}$. As a result, waves break over the planar beach starting at $x$ = 4~m shortly after they are generated at $x$ = 1~m. Wave breaking is initiated at the center of the short crest, spreading alongshore in both directions, towards the nodal lines. A series of isolated waves (i.e., individual waves with a higher wave height) are then generated at the crest ends. The generation of isolated waves is attributed to two factors\cite{wei2017short}. The major factor is the superposition of crest ends of intersecting waves cross the nodal line, resulting in an increase of local wave height at that location. A secondary factor, which takes place only after the development of rip currents, is wave-current interaction. Basically the rip current near the nodal line opposes and further slows down the onshore propagation of the crest end, resulting in local wave amplification and separation of isolated waves from the crest head. Moreover, wave amplitude diffraction at isolated breakers leads to an increase in the alongshore wave number in the inner surf zone.
It should be pointed out that isolated breakers are also observed with wave height $H$ = 0.2~m in Fig.~\ref{fig:eta}(a)--(c), however, they are smaller than those with $H$ = 0.3~m in Fig.~\ref{fig:eta}(d). 


\begin{figure*}[!t]
\centering
  \includegraphics[width=0.9\textwidth]{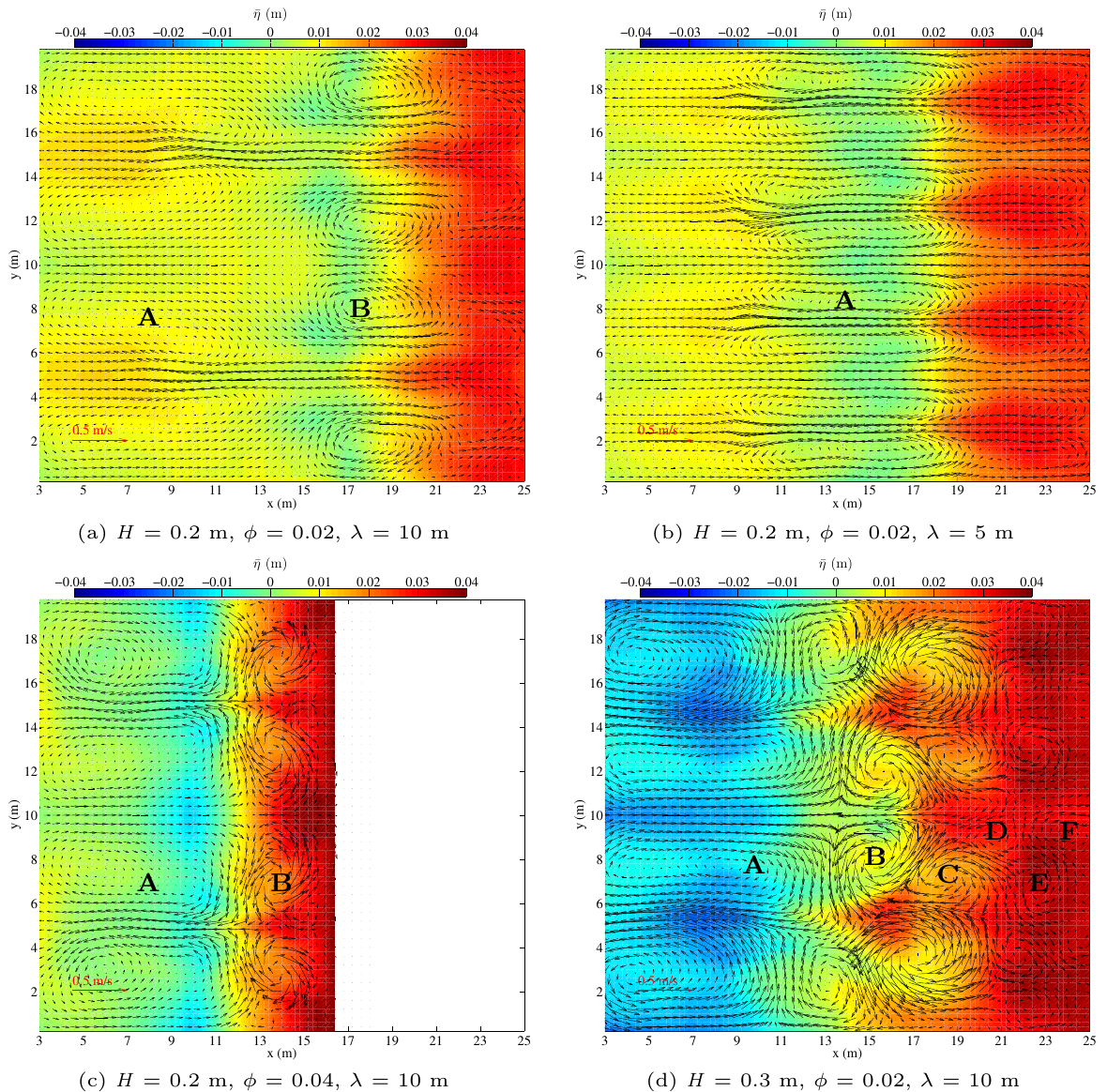}
\caption{The time- and depth-averaged current field and the mean water level of short-crested wave breaking over a beach among different numerical experiments. $H$ indicates the wave height; $\phi$ indicates the beach slope; $\lambda$ indicates the rip current spacing. Alphabet letters $A$ -- $F$ indicate circulation cells cross the wave basin. The location of nodal lines can be found in Fig.~\ref{fig:eta}.}
 \label{fig:eta_bar}
\end{figure*}
\section{Current Field} \label{current_field}

\begin{figure*}[!t]
\centering
  \includegraphics[width=0.9\textwidth]{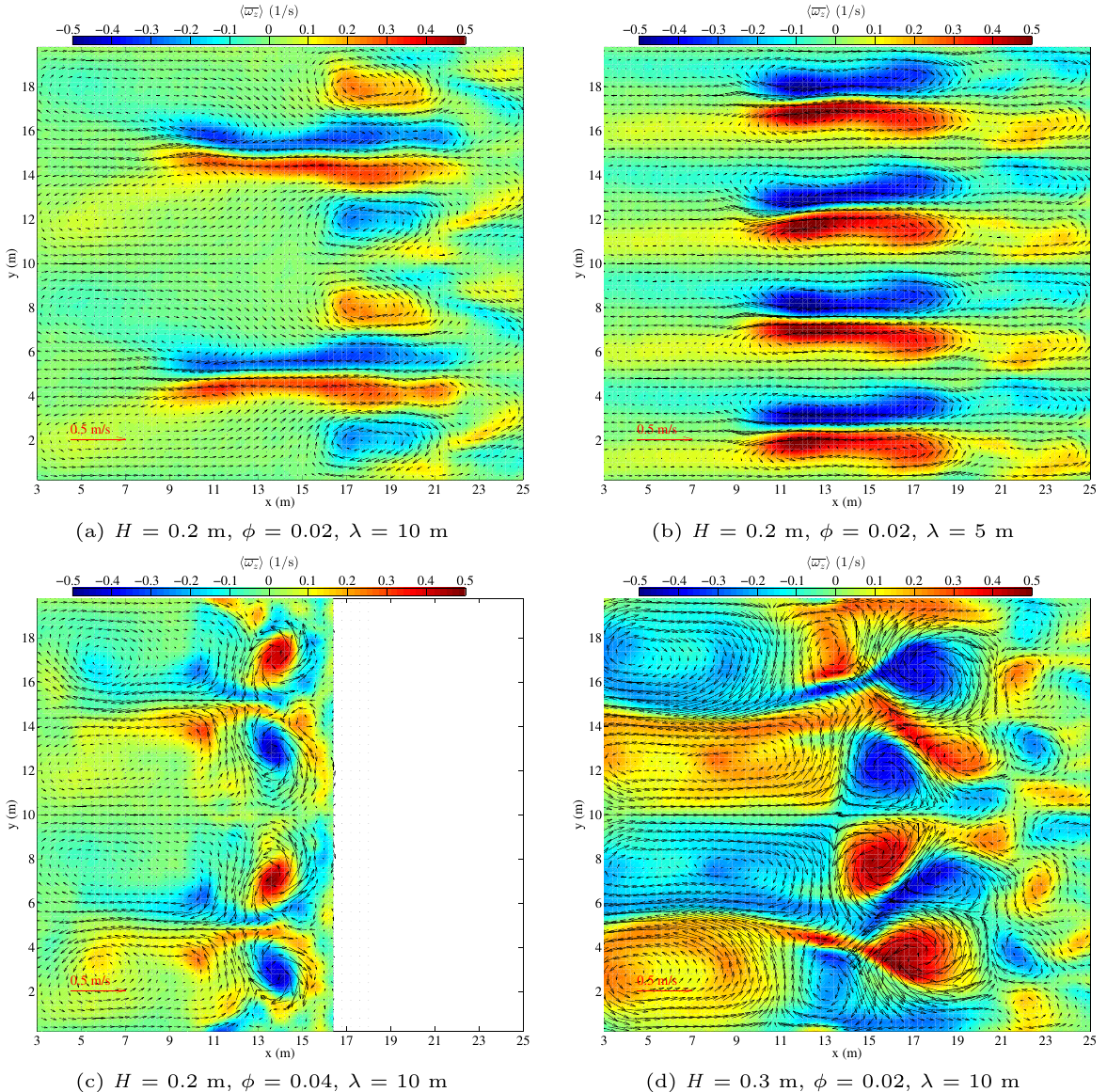}
\caption{The time- and depth-averaged vertical vorticity field and the associated time- and depth-averaged current field of short-crested wave breaking over a beach among different numerical experiments. $H$ indicates the wave height; $\phi$ indicates the beach slope; $\lambda$ indicates the rip current spacing. The location of nodal lines can be found in Fig.~\ref{fig:eta}.}
 \label{fig:omega}
\end{figure*}
\subsection{Time-averaged Vertical Current Field}

The time-averaged velocity at any point is defined by
\begin{equation}
\overline{ \vec{u} (x, y, z)} =\frac{1}{n_wT}\int_{t_0}^{t_0+n_wT}\vec{u}(x, y, z, t)\mathrm d t \label{eq:u_t}
\end{equation}
where $t_0 = 40 T$ is the start time of the measurement; the total number of waves sampled is $n_w$ = 10; and the SPH particle averaged velocity $\vec{u}$ has three components $(u , v, w)$.

Fig.~\ref{fig:vertical_current} shows the vertical distribution of the time-averaged cross-shore velocity $ \overline{ u (x, y, z)} $ at the nodal line (i.e., the location of rip current) for all cases. It has been seen in Fig.~\ref{fig:eta} that short-crested waves with $H$ = 0.2~m break over the planar beach. As a result, the generated rip currents are confined in a region above the planar beach, as shown in Fig.~\ref{fig:vertical_current}(a)--(c). However, a stronger rip current is observed when the rip current spacing is narrower (Fig.~\ref{fig:vertical_current}(b)) and when the beach is steeper (Fig.~\ref{fig:vertical_current}(c)). When the incident wave height is increased to be $H$ = 0.3~m, waves break near the beach toe at $x$ = 4~m, resulting in a stronger rip current as seen in Fig.~\ref{fig:vertical_current}(d). Theoretically speaking, there should be no wave crossing the nodal line since intersecting waves cancel each other along that line based on the linear wave theory, which is true before waves break (e.g., $x < $ 8~m in Fig.~\ref{fig:vertical_current}(a) and $x <$ 6~m in Fig.~\ref{fig:vertical_current}(c)). However, once the short-crested waves break over the beach, the phenomenon of wave crossing the nodal line can be observed in all subplots of Fig.~\ref{fig:vertical_current}. The wave height is determined by measuring the vertical distance between two horizontal dashed lines near the mean water level. It is seen that cases with a steeper beach (Fig.~\ref{fig:vertical_current}(c)) and a larger incident wave height (Fig.~\ref{fig:vertical_current}(d)) have a larger wave than the $Baseline$ case (Fig.~\ref{fig:vertical_current}(a)). This nonlinear wave phenomenon is attributed to: (1) the nonlinear interaction between intersecting waves cross the nodal line and (2) wave-current interaction, as explained in Wei et al.\cite{wei2017short} and manifested by the observation of isolated waves in Fig.~\ref{fig:eta}.


\subsection{Time- and Depth-averaged Current Field}
Next we analyze the mean water level and the current field over the basin. The mean water level is obtained by
\begin{equation}
\overline{\eta(x, y)} = \frac{1}{n_wT}\int_{t_0}^{t_0+n_wT} \eta(x,y) \mathrm d t
\end{equation}
and the time- and depth-averaged velocity is defined by
\begin{equation}
\overline{\langle \vec{u} (x, y)\rangle}=\frac{1}{z_0-z_b}\frac{1}{n_wT}\int_{z_b}^{z_0}\int_{t_0}^{t_0+n_wT}\vec{u}(x, y, z, t)\mathrm d z \mathrm d t \label{eq:u_td}
\end{equation}
where $z_b$ is the bottom of the basin, and $z_0$ is the mean water level. 

Fig.~\ref{fig:eta_bar} shows the time- and depth-averaged current field and the mean water level of short-crested wave breaking over a beach among different numerical experiments. We first analyze the wave setup over the wave basin. Due to wave shoaling, there is a wave setdown region (i.e., the mean water level is below zero) observed in all cases in Fig.~\ref{fig:eta_bar} (e.g., around $x$ = 17~m in Fig.~\ref{fig:eta_bar}(a), $x$ = 16~m in Fig.~\ref{fig:eta_bar}(b), $x$ = 10~m in Fig.~\ref{fig:eta_bar}(c), and $x < $ 10~m in Fig.~\ref{fig:eta_bar}(d)). After the wave breaks, the decrease of radiation stress is balanced by wave setup nearshore. It is seen that the wave setup is higher along the nodal line in all cases, and it drives the formation of rip currents. However, distributions of wave setup are different among four numerical experiments due to different breaking wave patterns, as seen in Fig.~\ref{fig:eta}. 

The current field in Fig.~\ref{fig:eta_bar} shows the existence of nearshore circulations inside the wave basin. For the \textit{Baseline} case in Fig.~\ref{fig:eta_bar}(a), there are two circulation cells. The offshore cell $A$ is between the nodal line $y$ = 5~m and the antinodal line $y$ = 10~m, and it is very weak. The other cell $B$ is driven by the wave setup near $y$ = 10~m nearshore, and it flows against the wave setup near the nodal line $y$ = 5~m. For \textit{Case} $\bf{b}$ in Fig.~\ref{fig:eta_bar}(b) with a larger incident wave angle, only one circulation cell is formed between the nodal line and the antinodal line, and it is likely that there is no alongshore spacing available to create the other cell $B$, as observed in Fig.~\ref{fig:eta_bar}(a). The single cell pattern in Fig.~\ref{fig:eta_bar}(b) is very similar to the one observed by Dalrymple\cite{dalrymple1975mechanism} in the laboratory. For \textit{Case} $\bf{c}$ in Fig.~\ref{fig:eta_bar}(c) with a steeper beach, there are two circulation cells. But both cells are stronger than those in the \textit{Baseline} case in Fig.~\ref{fig:eta_bar}(a). Furthermore, the secondary cell $B$ is mainly limited in the wave setup region and it also flows against the wave setup. \textit{Case} $\bf{d}$ in Fig.~\ref{fig:eta_bar}(d) has a larger incident wave height, resulting in a more complicated nearshore circulation pattern. It also has a primary circulation cell $A$ offshore as observed in Fig.~\ref{fig:eta_bar}(a) and (c), but it has more secondary circulation cells nearshore.

\section{Vertical Vorticity Field}\label{circulation}
The vertical vorticity that describes the horizontal rotation of flows is defined by
\begin{equation}
\omega_z = \frac{\partial u_y}{\partial x} -\frac{\partial u_x}{\partial y} \label{eq:omega}
\end{equation}
Furthermore, the time-averaged vertical vorticity and the time- and depth-averaged vertical vorticity can be computed by Eqs.~(\ref{eq:u_t}) and (\ref{eq:u_td}) with Eq.~(\ref{eq:omega}) as the input, respectively. 

Fig.~\ref{fig:omega} shows the time- and depth-averaged vertical vorticity field and the associated time- and depth-averaged current field of short-crested wave breaking over a beach among different numerical experiments. It is seen that the flow rotation direction is consistent with the sign of the vertical vorticity (e.g., a positive vertical vorticity indicates an anticlockwise circulation cell). Furthermore, Fig.~\ref{fig:omega} shows that rip current is formed in a region with opposite signed vertical vorticity (e.g., around $y$ = 5 and 10~m in Fig.~\ref{fig:omega}(a), and $y$ = 2.5, 7.5, 12.5, and 17.5~m in Fig.~\ref{fig:omega}(b)), as previously noticed by Johnson and Pattiaratchi\cite{johnson2006boussinesq}. In Wei et al.\cite{wei2017short}, we concluded that there are 3D vortex structures generated inside the wave basin after examining the nearshore circulation pattern, the time-averaged vertical vorticity field, and the vertical variation of vertical vorticity over the water column. We further suggested that the strong vortex motion drives a complex current pattern, such as current against the wave setup in Fig.~\ref{fig:eta_bar}(d). It is likely that our previous findings are still valid to explain the complicated current pattern as seen in Fig.~\ref{fig:eta_bar}(a) and (c).

\section{Conclusions}\label{conclusions}
In this study we conducted a series of numerical experiments to investigate short-crested waves in the surf zone with GPUSPH. We analyzed the breaking wave profile, the current field, and the vertical vorticity field under short-crested wave breaking over a planar beach by considering different parameters including beach slope, incident wave height, and incident wave angle. 
Our findings include:
  \begin{itemize}
\item Short-crested wave breaking over a steeper beach generates stronger rip currents, and larger circulation cells in front of the beach.
  \item Intersecting wave trains with a larger incident wave height drive a more complicated short-crested wave field including isolated breakers and wave amplitude diffraction.
  \item Nearshore circulation induced by short-crested wave breaking is greatly influenced by the incident wave angle (or the rip current spacing). There is no secondary circulation cell between the nodal line and the antinodal line if the rip current spacing is narrow. However, there are multiple secondary circulation cells observed when the rip current spacing is relatively large.

  \end{itemize}
It should be pointed out that the above observations were based on short-crested waves with two major wave components, future work is needed to verify these findings by examining nearshore waves in reality. Furthermore, a relatively coarse SPH particle size is used in this work, a finer SPH particle size is needed to better resolve the breaking wave field and the associated nearshore wave setup.

\section*{Acknowledgment}
The authors acknowledge the support from the Office of Naval Research, Littoral Geosciences and Optics Program. The authors also acknowledge the ATHOS Consortium and its member organizations for their contributions to the GPUSPH code. The numerical simulations were carried out at the Maryland Advanced Research Computing Center.



%



\end{document}